# Seasonal and Diurnal Variability of Atmospheric Pressure in Jezero Crater, Mars, from MEDA Measurements on the Perseverance Rover

Julio Carlos Bertua Marasca, Homer Dávila Gutiérrez, and Josué Ismael Mosquera Hadatty

Master's Degree in Astrophysics and Astronomy, Universidad Internacional de La Rioja (UNIR), Spain
Corresponding author: cosmos@skycr.org

## Abstract

Statistical analysis of atmospheric pressure measurements within Jezero Crater, recorded by the MEDA meteorological station aboard NASA's *Perseverance* rover, reveals variability consistent with known climatic patterns on Mars. Seasonal changes, axial tilt, and local geomorphological and atmospheric factors contribute to this variability. Notably, during Sol 504 an increase in atmospheric pressure was observed, despite this period coinciding with the onset of northern winter. The effect of airborne dust on pressure is significant and often reinforced by thermal tides, which cause both daily and monthly oscillations in the atmospheric pressure.

## Introduction

This study presents an analysis of the prevailing climatic conditions within Jezero Crater on Mars. The work focuses specifically on barometric conditions, providing a localized meteorological overview of this site. The data analyzed were obtained from the MEDA (Mars Environmental Dynamics Analyzer) weather station aboard NASA's *Perseverance* rover, which successfully landed in Jezero Crater in February 2021.

The study examines barometric records from Sols 182, 361, 504, and 658, representing a limited but representative sample of atmospheric pressure variability across the four Martian seasons. The results are also compared with data obtained by the *Curiosity* rover, which operates in the southern hemisphere near the planet's equatorial region.

This approach allows us to contrast meteorological behavior between two distinct sites under different geomorphological and latitudinal conditions. Although both rovers are located in relatively low-latitude regions, Jezero Crater lies slightly north of the equator, while Gale Crater is situated closer to the southern tropics. This positional difference provides valuable insight into how solar insolation, $CO_2$ condensation–sublimation cycles, and local topography influence pressure patterns on Mars.

## Materials and Methods

Data used in this study were obtained from NASA's Planetary Data System (PDS), managed by the Jet Propulsion Laboratory (JPL). Specifically, atmospheric pressure data were retrieved from the MEDA (Mars Environmental Dynamics Analyzer) instrument onboard the *Perseverance* rover.

The data were processed and analyzed using Python 3.14. Four pressure profiles were generated, corresponding to Sols 182, 361, 504, and 658, which represent four different seasonal periods on Mars. Because certain measurement intervals contained missing data ("gaps"), an averaging process was applied to every five-minute group of readings to ensure temporal consistency.

Although a more detailed study of barometric variations in Jezero Crater could be performed with the full MEDA dataset, the present analysis should be interpreted as a local characterization rather than a global representation of the Martian atmosphere. This limitation arises because the rover's measurements are influenced by local geomorphological conditions, elevation, and latitude.

Jezero Crater lies near the equatorial region of Mars (approximately 18°33′ N; Wimmer-Schweingruber et al., 2020). Therefore, its pressure behavior differs from that of higher-latitude regions, which are strongly affected by the seasonal condensation and sublimation of $CO_2$ at the poles (Mischna, 2020).

**General Overview**

Barometric, or atmospheric, pressure represents the force per unit area exerted by the column of gas in a planetary atmosphere upon the surface of the planet. In the case of Mars, the mean surface atmospheric pressure is only 6.1 mbar (Centro de Astrobiología, CSIC–INTA, 2017), substantially lower than that of Earth, whose most precise sea-level measurement reaches 1013.25 mbar, equivalent to 1013.25 hPa (NOAA, 2023).

The following analysis presents the in-situ barometric variations recorded within Jezero Crater by the MEDA (Mars Environmental Dynamics Analyzer) instrument aboard NASA's *Perseverance* rover. Pressure measurements corresponding to Sols 182, 361, 504, and 658 were selected as representative samples of the atmospheric behavior during the four Martian seasons. These data provide direct evidence of temporal variability in local atmospheric pressure, influenced by both seasonal cycles and site-specific geomorphological conditions.

| PRESIÓN ATMOSFÉRICA EN EL CRÁTER JEZERO, MARTE | | | | |
|---|---|---|---|---|
| | SOL | | | |
| | 182 | 361 | 504 | 658 |
| Mediciones utilizables | 24093 | 64896 | 36939 | 69282 |
| Media | 715.66 | 653.11 | 784.65 | 726.18 |
| Desviación Estándar | 5.10 | 12.07 | 11.26 | 12.54 |
| % Desv/Media | 0.71 | 1.85 | 1.44 | 1.73 |
| Presión Mínima | 708.40 | 632.51 | 767.45 | 706.72 |
| Hora del Mínimo | 3:02 | 17:55 | 16:40 | 17:10 |
| Presión Máxima | 725.74 | 676.33 | 804.06 | 754.67 |
| Hora del Máximo | 7:20 | 7:57 | 21:47 | 7:45 |
| Amplitud | 17.34 | 43.82 | 36.61 | 48.33 |

Tabla no.1. Elaboración propia en base datos de NASA.

Table 1 summarizes the atmospheric pressure measurements within Jezero Crater, showing that the lowest mean values occur during the northern winter. This can be explained by the reduced solar irradiance at that time of year: around Sol 361, corresponding to the northern winter equinox, lower insolation favors the condensation of $CO_2$ at the poles, thereby decreasing the total mass of the atmospheric column.

An opposite but complementary phenomenon occurs near the northern winter solstice, when the mean pressure reaches its annual maximum. This increase is associated with the onset of the northern winter season, during which enhanced $CO_2$ sublimation and redistribution lead to a denser atmospheric column.

Mars has an axial tilt of approximately 25.2° (Jancke et al., 2004). Combined with the latitude of Jezero Crater (18°33′ N) (Wimmer-Schweingruber et al., 2020), these parameters yield a quasi-equatorial climate characterized by moderate seasonal variation and by the presence of thermal-tide harmonics that impart a distinctive oscillatory behavior to the pressure curve (Hernández et al., 2024).

Moreover, these variations are modulated by Mars's orbital configuration. Around Sol 504, the planet was near perihelion, receiving increased solar radiation that intensified atmospheric heating and expanded the gaseous column (Hansen et al., 2024). The influence of this radiative forcing is illustrated in Figure 1, adapted from Hansen et al. (2024), which demonstrates the significant role of orbital position in shaping pressure variability at Jezero Crater.

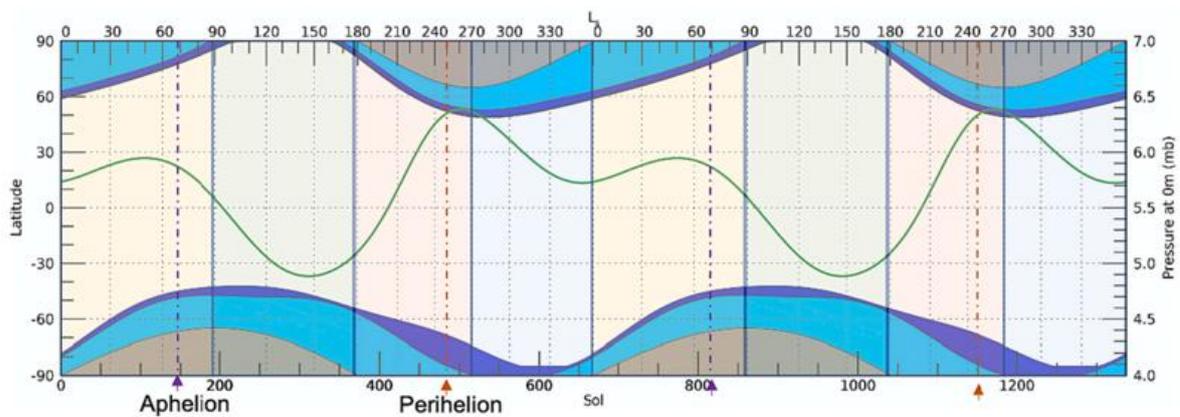

Figure 1. *The pressure cycle over two Martian years* (green curve, right axis) *exhibits two distinct pressure drops per year.* The first, more pronounced decline occurs during northern summer, when $CO_2$ condenses to form the seasonal frost cap in the southern hemisphere. The second, smaller decrease takes place at the end of southern summer, coinciding with the formation of the northern winter cap. The maximum atmospheric pressure aligns approximately with perihelion, when Mars receives its greatest solar irradiance. Colored shading denotes the northern hemisphere's spring (yellow), summer (green), autumn (light orange), and winter (blue). The gray shading represents the latitudinal extent of the polar night (left axis). Light and dark blue bands indicate the minimum and maximum latitudinal extents of the seasonal $CO_2$ frost, respectively—also referred to as the inner and outer crocus lines (Titus, 2005). The pressure curve is derived from *Tillman et al. (1993). Figure adapted from Hansen et al. (2024).*

Hernández et al. (2024) found that thermal-tide harmonics are present in the annual pressure variations of the Martian atmosphere, as illustrated in Figure 2.

These harmonics reflect the periodic modulation of surface pressure driven by solar heating and atmospheric dynamics throughout the Martian year.

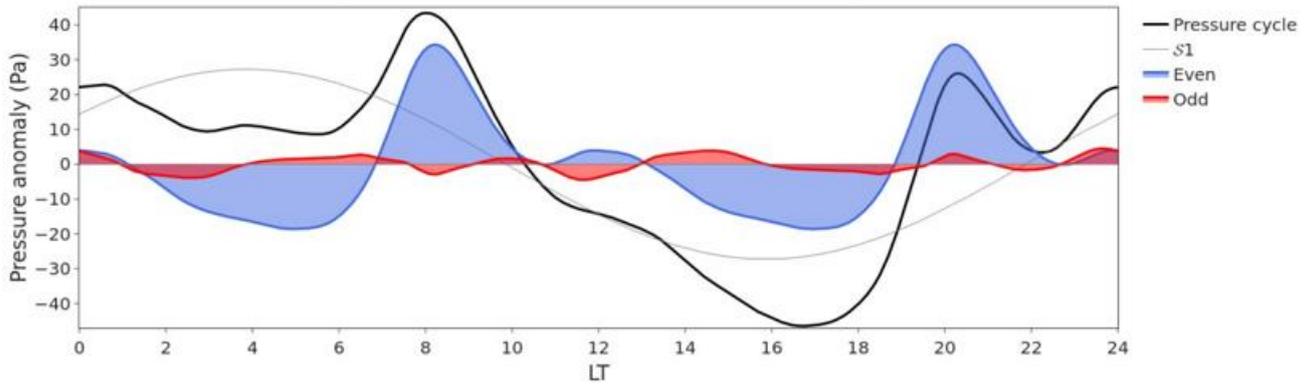

Figure 2. *Differences between even and odd harmonics of atmospheric pressure derived using the Fast Fourier Transform (FFT) method.* The analysis reveals distinct harmonic components associated with thermal tides throughout the Martian year. *After Hernández et al. (2024).*

Summer Solstice

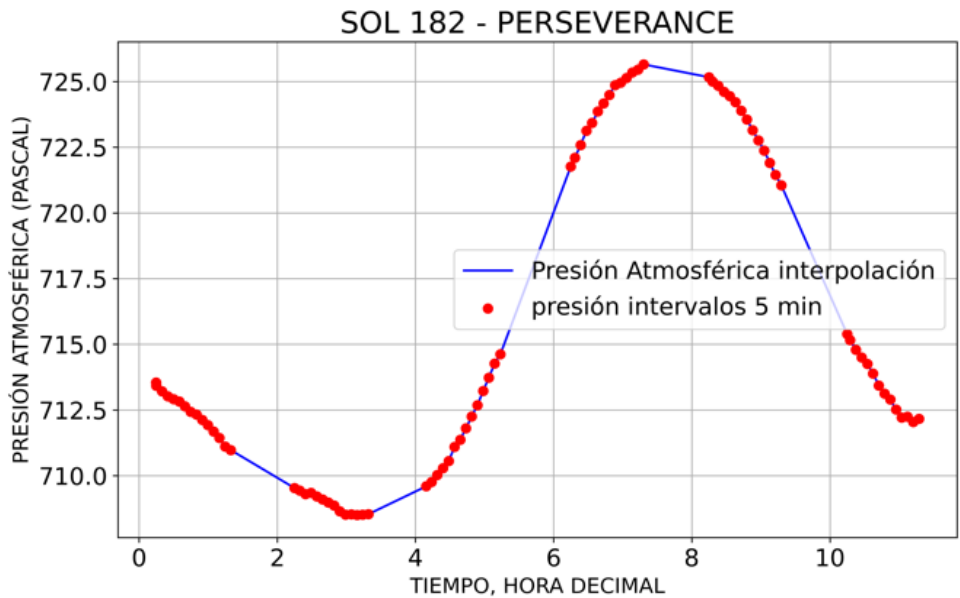

Figure 3. *Atmospheric pressure from MEDA data is shown in red, while interpolated values for missing intervals are displayed in blue.* During Sol 182, corresponding to the northern summer solstice, the pressure exhibits a quasi-Gaussian diurnal profile, peaking during the hours of greatest solar irradiance. The limited number of usable measurements justifies the truncated coverage of the full Martian day. *Figure produced by the authors using MEDA data from NASA, processed in Python version 3.12.*

Winter Equinox

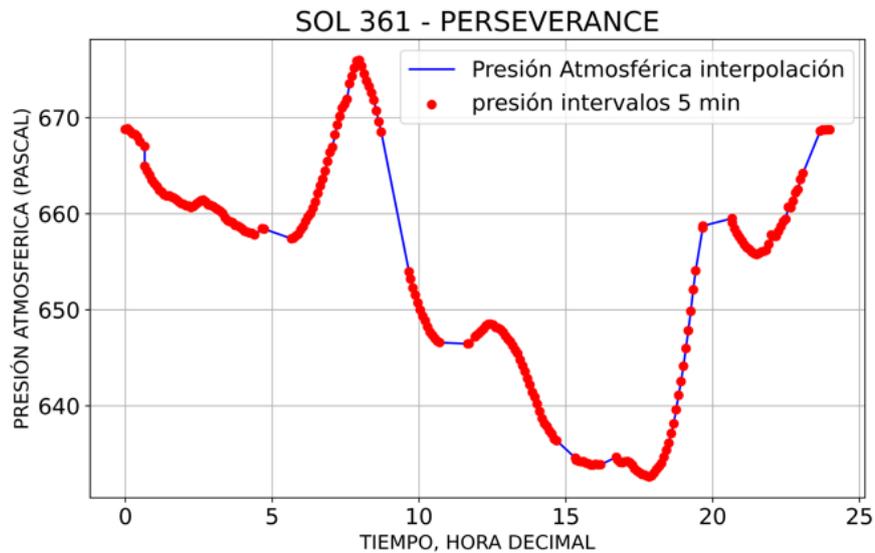

Figure 4. *Atmospheric pressure from MEDA data (red) and interpolated values for missing data (blue).* The drop in pressure is an expected feature during the onset of northern winter, as shown here for the northern winter equinox, when seasonal $CO_2$ deposition reaches its maximum extent. *Figure produced by the authors using MEDA data from NASA, processed in Python version 3.12.*

Winter Solstice

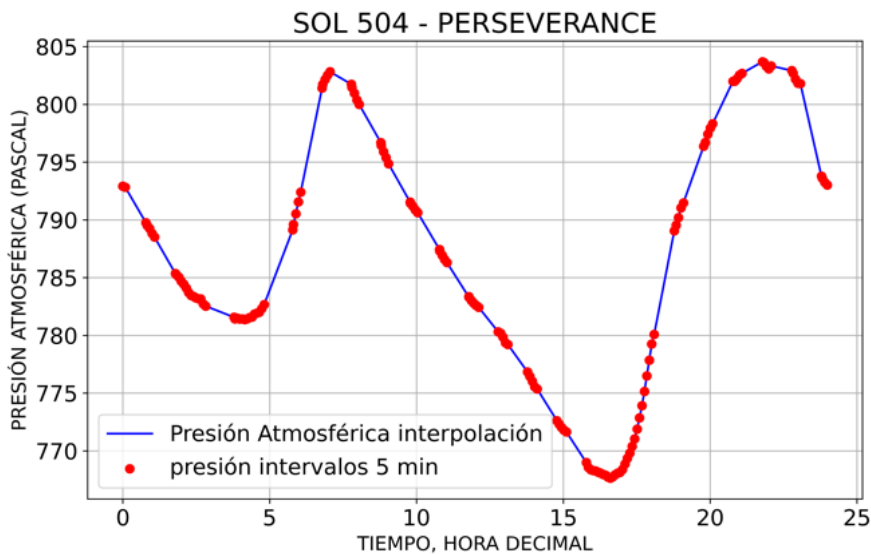

Figure 5. *Atmospheric pressure derived from MEDA data is shown in red, while interpolated values for missing intervals are displayed in blue.* One of the most intriguing features in the barometric variability observed at Jezero Crater occurs during the northern winter solstice. As previously discussed, this phenomenon results from multiple interacting causes, including orbital dynamics, thermal tides, and local geomorphological effects, all of which modulate the pressure amplitude. *Figure produced by the authors using MEDA data from NASA, processed in Python version 3.12.*

Spring Equinox

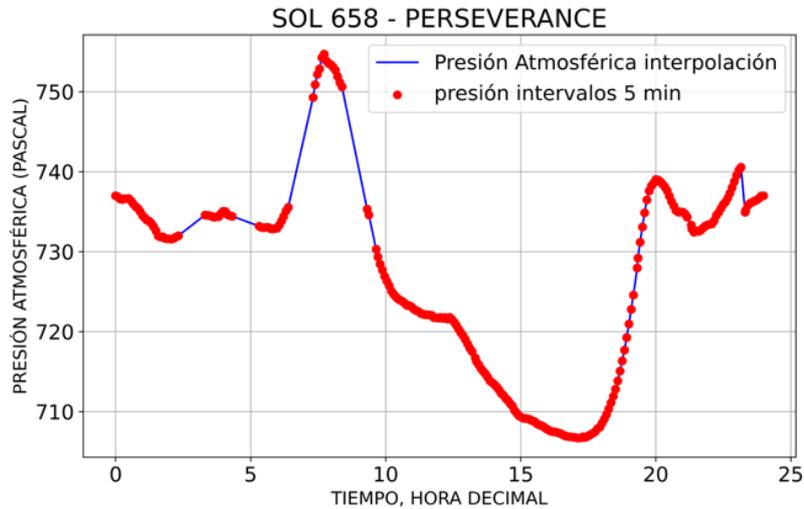

Figure 6. *Atmospheric pressure derived from MEDA data is shown in red, while interpolated values for missing data are displayed in blue.* During the spring equinox, the overall pressure trend resembles that of the winter equinox, though with higher mean values. This increase occurs because there is neither $CO_2$ condensation nor sublimation during this transitional season. *Figure produced by the authors using MEDA data from NASA, processed in Python version 3.12.*

To fully understand the pressure variability observed in MEDA data, it is essential to consider that Mars is frequently affected by dust storms and by localized vortices known as dust devils, both of which generate turbulent flows that perturb local pressure fields, as illustrated in Figure 7.

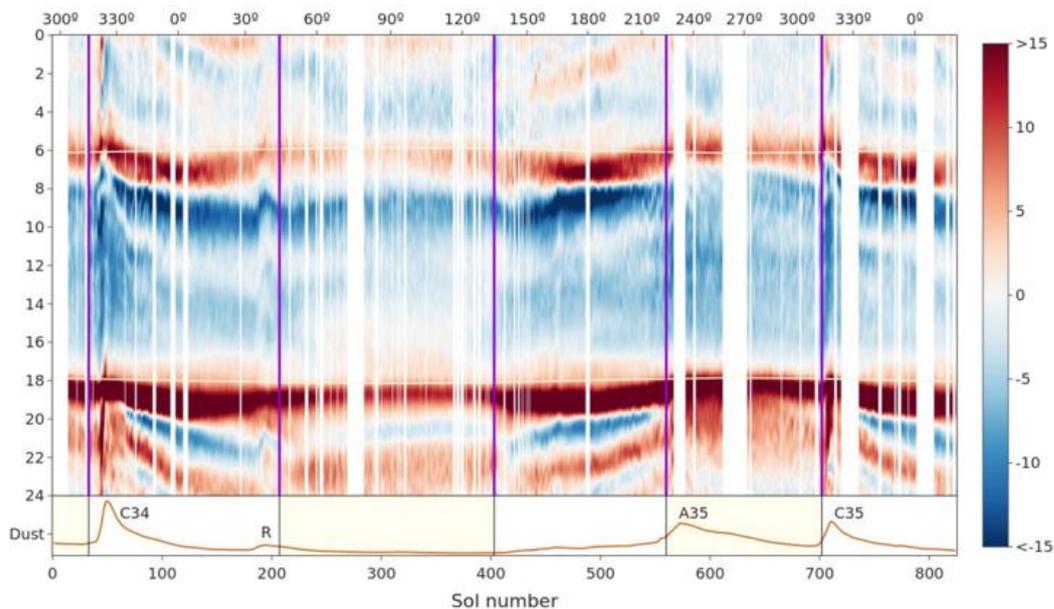

Figure 7. *Climatology of diurnal pressure variation (dP/dt) in units of Pa h$^{-1}$.* High-frequency signals (with periods shorter than 3,700 s) were removed from the original time series prior to differentiation. Sunrise and sunset are indicated by white curves around 06:00 and 18:00 LT (local true solar time). The global dust content, calculated from the climatologies of Montabone et al. (2015; 2020), is displayed in the lower panel, where individual dust events are also identified. Seasons are delimited by vertical lines, and solstices are shaded in pale yellow in the lower panel (dust). *Figure adapted from Hernández et al. (2024).*

# Conclusions

An analysis of the variability of atmospheric pressure in Jezero Crater, compared with corresponding data from the Curiosity rover in the southern hemisphere, shows that Martian surface pressure is governed by several interrelated factors and phenomena:

1. Temperature Variations
   Because Mars's atmosphere is much thinner than Earth's, equatorial and mid-latitude regions warm rapidly during the day but also cool quickly at night. This produces significant diurnal swings in atmospheric pressure. At night, rapid cooling increases air density, enhancing the weight of the atmospheric column. In addition, Mars experiences pronounced seasonal cycles driven by its axial tilt, causing $CO_2$ to condense rapidly at the poles during winter, which lowers global pressure. Conversely, in summer, the $CO_2$ frost sublimates directly to gas rather than melting into liquid form as on Earth, thereby raising the global atmospheric pressure. All of this occurs while Mars continues to lose atmospheric mass over geological timescales.
2. Dust Effects
   Mars is repeatedly affected by planet-wide dust storms, which can envelop the entire planet within weeks. These global events substantially influence atmospheric pressure by altering radiative balance, creating a localized greenhouse effect that warms the atmosphere and modifies both air density and pressure.
3. Thermal Tides
   Thermal tides are well documented and represent one of the main causes of regional pressure oscillations. This phenomenon appears in the figures as daytime pressure decreases and nighttime increases (as observed in Figures 4 and 6). Certain nocturnal pressure enhancements remain unexplained and may require additional investigation.
4. Local Geomorphology
   As noted earlier, local topography strongly modulates pressure variability at both Jezero and Gale Craters. Wind currents interacting with crater walls or adjacent plains produce localized perturbations in pressure. Consequently, the atmospheric conditions inside a crater—or atop a nearby plateau—cannot be generalized to represent the global Martian atmosphere, given the planet's highly variable terrain and dynamic environmental processes.